\documentclass[letterpaper,twocolumn,english,superscriptaddress]{revtex4}
\usepackage{mathptmx}
\usepackage[T1]{fontenc}
\usepackage[latin9]{inputenc}
\usepackage{amsmath}
\usepackage{graphicx}
\usepackage{amssymb}
\usepackage{esint}

\makeatletter

\@ifundefined{textcolor}{}
{%
 \definecolor{BLACK}{gray}{0}
 \definecolor{WHITE}{gray}{1}
 \definecolor{RED}{rgb}{1,0,0}
 \definecolor{GREEN}{rgb}{0,1,0}
 \definecolor{BLUE}{rgb}{0,0,1}
 \definecolor{CYAN}{cmyk}{1,0,0,0}
 \definecolor{MAGENTA}{cmyk}{0,1,0,0}
 \definecolor{YELLOW}{cmyk}{0,0,1,0}
 }

\usepackage{mathptmx}

\usepackage{babel}

\usepackage{babel}

\makeatother

\usepackage{babel}

\begin{document}

\title{Coherence of a qubit stored in Zeeman levels of a single optically
trapped atom}

\author{Wenjamin Rosenfeld}

\email[corresponding author: ]{wenjamin.rosenfeld@physik.uni-muenchen.de}

\affiliation{Fakultät für Physik, Ludwig-Maximilians-Universität München, D-80799
München, Germany}

\affiliation{Max-Planck-Institut für Quantenoptik, D-85748 Garching, Germany}

\author{Jürgen Volz}

\altaffiliation[present address: ]{ Laboratoire Kastler Brossel de l'E.N.S. 24, rue Lhomond F-75005 Paris, France}

\affiliation{Fakultät für Physik, Ludwig-Maximilians-Universität München, D-80799
München, Germany}

\author{Markus Weber}

\affiliation{Fakultät für Physik, Ludwig-Maximilians-Universität München, D-80799
München, Germany}

\author{Harald Weinfurter}

\affiliation{Fakultät für Physik, Ludwig-Maximilians-Universität München, D-80799
München, Germany}

\affiliation{Max-Planck-Institut für Quantenoptik, D-85748 Garching, Germany}
\begin{abstract}
We experimentally investigate the coherence properties of a qubit
stored in the Zeeman substates of the $5^{2}S_{1/2},\: F=1$ hyperfine
ground level of a single optically trapped $^{87}$Rb atom. Larmor
precession of a single atomic spin-1 system is observed by preparing
the atom in a defined initial spin-state and then measuring the resulting
state after a programmable period of free evolution. Additionally,
by performing quantum state tomography, maximum knowledge about the
spin coherence is gathered. By using an active magnetic field stabilization
and without application of a magnetic guiding field we achieve transverse
and longitudinal dephasing times of $T_{2}^{*}=75..150\,\mathrm{\mu s}$
and $T_{1}>0.5\,\mathrm{ms}$ respectively. We derive the light-shift
distribution of a single atom in the approximately harmonic potential
of a dipole trap and show that the measured atomic spin coherence
is limited mainly by residual position- and state-dependent effects
in the optical trapping potential. The improved understanding enables
longer coherence times, an important prerequisite for future applications
in long-distance quantum communication and computation with atoms
in optical lattices or for a loophole-free test of Bell's inequality.
\end{abstract}


\maketitle

\section{Introduction}

Quantum memories for the storage and retrieval of quantum information
play an outstanding role in future applications of quantum communication,
such as quantum networks and the quantum repeater \cite{Briegel98}.
There, ground states of trapped atoms or ions, are ideal candidates,
as the interaction with the environment is weak and can be controlled
with high accuracy. Although in such systems coherence times of the
order of several seconds have been observed \cite{Davidson95,Kuhr05,Ozeri05},
storage and retrieval of single quantum excitations was shown to reach
maximum times of only few $100\,\mathrm{\mu s}$ to $\mathrm{ms}$
\cite{Choi08,Rosenfeld08,Zhao09,Dudin09,Specht11}, in which storage
of a complete polarization qubit state represents a greater challenge.
In order to further prolong the quantum storage time a detailed understanding
of dephasing and decoherence processes is indispensable.

Optical dipole traps \cite{Grimm00} nowadays are a well-established
tool for the controlled manipulation of internal and external quantum
states of neutral atoms \cite{Kuhr05,Nelson07,Bloch08,Rosenfeld08,Dudin09,Foerster09,Bakr09,Wilk10,Isenhower10}.
Such traps provide almost ideal conservative trapping potentials combined
with low heating rates, resulting in long atomic coherence times.
However, two kind of effects significantly limit the achievable coherence
time: (i) Off-resonant spontaneous Raman scattering from the dipole
laser beam entangles the qubit of freedom with some degree of freedom
of a single scattered photon \cite{Blinov04,Matsukevich05,APE06}.
This kind of light-matter interaction leads to decoherence in the
most general sense as the system under investigation (atomic memory
qubit) gets entangled with the environment (scattered photon). (ii)
In addition, if the trapped atom is not in the vibrational ground
state, the thermal motion together with residual state-dependent effects
of the optical trapping potential will lead to dephasing of an initial
atomic spin state. Although the temporal spin-evolution in an individual
experimental realization is strictly coherent, the observed ensemble
average over many experimental runs may have a large scatter due to
different initial conditions.

In our experiment quantum information is stored in the Zeeman sublevels
of the $5^{2}S_{1/2},\: F=1$ hyperfine ground level of a single $^{87}$Rb
atom, localized in an optical dipole trap \cite{Weber06}. For a variety
of applications in long-distance quantum communication, e.g., the
generation of long-distance atom-photon \cite{APE06,Rosenfeld08}
and atom-atom entanglement, as well as the closely related task of
the remote preparation of an atomic quantum memory \cite{Rosenfeld07},
only a spin-1/2 subspace of the $5^{2}S_{1/2},\: F=1$ hyperfine ground
level is addressed. More precisely, the atomic memory qubit is encoded
in the $\left|F=1,\, m_{F}=-1\right\rangle $ and $\left|F=1,\, m_{F}=+1\right\rangle $
Zeeman sublevels. The remaining third sublevel $\left|F=1,\, m_{F}=0\right\rangle $
is not directly used for qubit storage, however, the coherent evolution
of the total angular momentum $F=1$ in a magnetic field (Larmor-precession)
can lead to its population, thereby reducing the fidelity of the stored
state. In order to extract information how a stored quantum state
becomes mixed and also to distinguish coherent from incoherent processes,
the temporal evolution of the full spin-1 density matrix of the $5^{2}S_{1/2},\: F=1$
hyperfine ground level has to be investigated.

In this contribution we analyze decoherence and dephasing mechanisms
and their relevance for the storage of quantum information in single
optically trapped $^{87}$Rb atoms. In detail, in Sec. \ref{sec:Theory}
we develop a model accounting for state-dependent effects of the optical
trapping potential. In order to achieve spin-coherence times of several
$100\,\mathrm{\mu s}$ we implement an active magnetic field stabilization
(see Sec. \ref{sec:Experiment}). Applying partial quantum state tomography
of the total angular momentum state $5^{2}S_{1/2},\: F=1$ we then
investigate in detail dephasing and decoherence mechanisms. Finally,
in Sec. \ref{sec:Conclusion} we summarize our major findings and
give an outlook how longer spin coherence times could be reached.

\section{Theoretical analysis of decoherence and dephasing mechanisms\label{sec:Theory}}

\subsection{Decoherence due to spontaneous Raman scattering}

The most influential scattering process occurs due to interaction
of the atom with the light of the dipole trap. In our case the dipole
trap is generated by a single, sharply focused laser beam at a wavelength
of $\lambda=856\,\mathrm{nm}$. Despite of the large detuning to the
first dipole allowed transitions $5^{2}S_{1/2}\rightarrow5^{2}P_{1/2}$
($\lambda=795\,\mathrm{nm}$) and $5^{2}S_{1/2}\rightarrow5^{2}P_{3/2}$
($\lambda=780\,\mathrm{nm}$) in $^{87}$Rb, there is still a finite
probability to spontaneously scatter light from the dipole laser beam.
This scattering process consists of two important parts. Elastic (Rayleigh)
scattering occurs when the atom returns to the same state after emission
of a photon. In our case this happens at a rate of $17.7\,\mathrm{Hz}$
\cite{Rosenfeld08_PhD}. When the final and initial states are different,
even though the states might be degenerate in energy, this process
is called spontaneous Raman scattering. As demonstrated previously
\cite{Blinov04,APE06,Matsukevich05}, it is this scattering process
which entangles, e.g., the polarization and/or the frequency of a
single scattered photon with the internal spin state of one or many
atoms. Obviously, only spontaneous Raman scattering will lead to spin
relaxation. Far from the atomic resonances $5^{2}S_{1/2}\rightarrow5^{2}P_{1/2}$
and $5^{2}S_{1/2}\rightarrow5^{2}P_{3/2}$ spontaneous Raman scattering
is strongly suppressed due to destructive interference of amplitudes
in the different excitation and decay channels \cite{Cline94}. Its
rate is given by \begin{equation}
\Gamma_{incoh}=\frac{3c^{2}\omega^{3}}{4\hbar}\left(\frac{\Gamma_{D}}{\omega_{D}^{3}}\right)^{2}\left|\frac{1}{\Delta_{D1}}-\frac{1}{\Delta_{D2}}\right|^{2}\cdot I,\label{eq:incoherentScattering}\end{equation}
where $\omega$ is the angular frequency of the dipole laser, $I$
its intensity, $\Delta_{D1}$ and $\Delta_{D2}$ its detuning with
respect to transitions to $5^{2}P_{1/2}$ and $5^{2}P_{3/2}$ levels,
$\Gamma_{D}$ is the total resonant scatter rate of the respective
$D1$ and $D2$ transition, and $\omega_{D}$ the angular frequency
of the corresponding atomic resonance. The rate for spontaneous Raman
scattering via higher-lying $n^{2}P_{1/2}$ and $n^{2}P_{3/2}$ levels
($n>5$) is negligible due to the large detuning of the dipole laser.
For our typical dipole trap parameters ($\lambda=856\,\mathrm{nm}$,
$I=1.56\cdot10^{9}\,\mathrm{W/m^{2}}$) the incoherent scatter rate
is $0.11\,\mathrm{Hz}$. Thus, the atomic spin will relax on a time
scale of about $10\,\mathrm{s}$, which is comparable to the lifetime
of captured atoms in the trap \cite{Weber06}. We conclude that in
for our experiments spin relaxation due to spontaneous Raman scattering
does not limit the coherence time.

\subsection{Dephasing of the atomic spin}

In contrast to the possibility that the atomic spin decoheres due
to entanglement with the environment, the coherent interaction with
fluctuating external magnetic fields $\vec{B}$ leads to dephasing
of stored quantum information. Additional spin-dephasing for trapped
atoms can also be caused by the thermal motion of the atom in a state-dependent
trapping potential (see Sec. \ref{sub:DistrLightShifts}).

The optical dipole trap is based on a spatially varying, light-induced
energy shift of the atomic ground levels. Assuming that all relevant
detunings are larger than the hyperfine ground state splitting we
obtain the light-shift for the Zeeman sublevels $\left|F=1,\, m_{F}\right\rangle $
of the $5^{2}S_{1/2}$ ground level of (see e.g. \cite{Grimm00})
\begin{equation}
\Delta E_{d}=-\frac{\pi c^{2}}{2}\frac{\Gamma_{D}}{\omega_{D}^{3}}\left(\frac{1-\mathcal{P}g_{F}m_{F}}{\Delta_{D1}}+\frac{2+\mathcal{P}g_{F}m_{F}}{\Delta_{D2}}\right)I.\label{eq:vectorLightShift}\end{equation}
For linear polarization of the trap light ($\mathcal{P}=0$) this
energy shift is equal for all magnetic sublevels $m_{F}$ of both
hyperfine ground levels $5^{2}S_{1/2},F=1$ and $5^{2}S_{1/2},F=2$
and therefore ideally suited as a state-insensitive trapping potential.
However, in the case of circular polarization ($\mathcal{P}=+1$ for
$\sigma^{+}$ and $\mathcal{P}=-1$ for $\sigma^{-}$) the shift becomes
state-dependent, lifting the degeneracy of the Zeeman sublevels. This
additional effect on the magnetic sublevels $m_{F}\ne0$ for circularly
polarized light is formally equivalent to a magnetic field $B_{\sigma}$
pointing along the propagation direction $z$ of the dipole laser
beam, and is called vector light shift (also Zeeman light shift) \cite{vectorLightShift}.
In contrast to an external magnetic field which can be considered
homogeneous over the microscopic volume of the optical dipole trap,
the vector light shift depends on the intensity and thus on the position
of the atom in the trap. For a typical temperature of the atom of
$100..150\,\mathrm{\mu K}$ the thermal motion will lead to a non-negligible
variation of light shifts, which significantly influence the dephasing
of stored quantum information. This dephasing mechanism will be analyzed
in the following.

\subsubsection{Coherent state evolution}

As a first step we calculate the coherent temporal evolution of the
spin-1 state $|\Psi(t)\rangle$ in a constant effective magnetic field
$B_{eff}$. The interaction Hamiltonian $\hat{H}_{eff}$ is given
by \begin{equation}
\hat{H}_{eff}=\vec{B}_{eff}\cdot\frac{\mu_{B}g_{F}}{\hbar}\hat{\vec{F}},\label{eq:fieldHamiltonian}\end{equation}
where $\hat{\vec{F}}$ is the operator of the corresponding angular
momentum $F=1$, $\mu_{B}=\hbar\cdot2\pi\cdot1.4\,\mathrm{MHz/G}$
is the Bohr magneton and $g_{F}=-1/2$ the Landé factor. The effective
magnetic field $\vec{B}_{eff}$ is given by \begin{equation}
\vec{B}_{eff}=\vec{B}+B_{\sigma}\cdot\vec{\mathrm{e}}_{z}=\vec{B}+\mathcal{P}\frac{1}{\mu_{B}}\frac{\pi c^{2}}{2}\frac{\Gamma_{D}}{\omega_{D}^{3}}\left(\frac{1}{\Delta_{D1}}-\frac{1}{\Delta_{D2}}\right)I_{\sigma}\cdot\vec{\mathrm{e}}_{z},\label{eq:effectiveField}\end{equation}
with $I_{\sigma}$ the intensity of the circularly polarized component
of the dipole trap beam. For convenience we set $\vec{B}_{eff}=B_{eff}(b_{x}\vec{\mathrm{e}}_{x}+b_{y}\vec{\mathrm{e}}_{y}+b_{z}\vec{\mathrm{e}}_{z})$,
where $b_{x}=\sqrt{1-b_{z}^{2}}\cos(\phi)$, $b_{y}=\sqrt{1-b_{z}^{2}}\sin(\phi)$.

For this field configuration we obtain the eigenstates $\left|\Phi_{-1}\right\rangle ,\,\left|\Phi_{0}\right\rangle ,\,\left|\Phi_{+1}\right\rangle $
of the interaction Hamiltonian $\hat{H}_{eff}$ \begin{align}
\left|\Phi_{+1}\right\rangle = & \left(\begin{array}{c}
\frac{1}{2}(1+b_{z})e^{-i\phi}\\
\frac{1}{\sqrt{2}}\sqrt{1-b_{z}^{2}}\\
\frac{1}{2}(1-b_{z})e^{i\phi}\end{array}\right),\nonumber \\
\left|\Phi_{0}\right\rangle = & \left(\begin{array}{c}
-\frac{1}{\sqrt{2}}\sqrt{1-b_{z}^{2}}e^{-i\phi}\\
b_{z}\\
\frac{1}{\sqrt{2}}\sqrt{1-b_{z}^{2}}e^{i\phi}\end{array}\right),\nonumber \\
\left|\Phi_{-1}\right\rangle = & \left(\begin{array}{c}
\frac{1}{2}(1-b_{z})e^{-i\phi}\\
-\frac{1}{\sqrt{2}}\sqrt{1-b_{z}^{2}}\\
\frac{1}{2}(1+b_{z})e^{i\phi}\end{array}\right),\label{ch4-eq:eigenstatesBeff}\end{align}
with corresponding eigenvalues $+\hbar\omega_{L}$, $0$, $-\hbar\omega_{L}$,
and Larmor frequency $\omega_{L}=\frac{1}{\hbar}\mu_{B}g_{F}B_{eff}$.
These are $m_{F}=+1,0,-1$ eigenstates with respect to the direction
of the effective magnetic field. Finally, for the time evolution of
an arbitrary $5^{2}S_{1/2},\, F=1$ state we obtain \begin{equation}
\left|\Psi(t)\right\rangle =c_{-1}\left|\Phi_{-1}\right\rangle e^{i\omega_{L}t}+c_{0}\left|\Phi_{0}\right\rangle +c_{+1}\left|\Phi_{+1}\right\rangle e^{-i\omega_{L}t},\label{eq:stateEvolution}\end{equation}
where the amplitudes are given by $c_{\pm1}=\left\langle \Phi_{\pm1}|\Psi(t=0)\right\rangle $,
$c_{0}=\left\langle \Phi_{0}|\Psi(t=0)\right\rangle $. Obviously,
the Larmor precession in an effective magnetic field $\vec{B}_{eff}$
is thus not necessarily limited to the qubit space $\{\left|1,-1\rangle,\right|1,+1\rangle\}$.

\subsubsection{Fluctuations of the magnetic fields\label{sub:FieldFluctuations}}

As a first step in the analysis of dephasing mechanisms one has to
know the magnitude of magnetic field fluctuations at the relevant
time scales. Typically, our experiments require preservation of the
atomic quantum state for several microseconds \cite{remark01}. In
this work we consider times up to $200\,\mathrm{\mu s}$, defining
two important frequency ranges. The first range contains frequencies,
where the magnetic field varies rapidly on the experimental time scale
($\Omega>2\pi\cdot2.5\,\mathrm{kHz}$). In the second range we have
$\Omega<2\pi\cdot2.5\,\mathrm{kHz}$, i.e., the field can be considered
constant within one experimental run, but will vary between different
runs. With the help of a magneto-resistive sensor (accessible frequency
range: $\mathrm{DC}-60\,\mathrm{kHz}$) we were able to quantify the
magnitude of magnetic fields at different frequencies, see Sec. \ref{sub:magneticFieldControl}.
We found that the strongest fluctuations were at low frequencies ($<200\,\mathrm{Hz}$)
while faster fluctuations were relatively small ($<0.3\,\mathrm{mG}$
rms within the bandwidth of $60\,\mathrm{kHz}$). Magnetic field fluctuations
at different frequencies affect the atomic state in different ways
as will be discussed below.

In the case where the field fluctuates rapidly on the experimental
time scale (and also rapidly compared to the Larmor frequency $\omega_{L}$),
the evolution of the atomic state will follow the average field $\bar{B}$
with small oscillations around the main trajectory. If the average
field $\bar{B}$ is constant, only those deviations will lead to dephasing.
The magnitude of these deviations can be estimated by solving the
time-dependent Schrödinger equation in a field modulated at a frequency
$\Omega\gg\omega_{L}$ and was found to drop with increasing modulation
frequency as $1/\Omega^{2}$ \cite{Rosenfeld08_PhD}. This can be
understood as the atomic spin, which rotates at a finite Larmor frequency,
can not follow the increasing frequency of the field fluctuations
which therefore average out. Thus, as the magnitude of rapid fluctuations
in our experiment is small ($<0.3\,\mathrm{mG}$) and due to the additional
$1/\Omega^{2}$ suppression, we conclude that the influence of rapid
oscillations is negligible, particularly when compared to the effect
due to the slowly varying field component.

For fluctuations of the magnetic field which are slow on the time
scale of a single experimental run, the field can be considered constant
and the atomic state will evolve according to Eq. (\ref{eq:stateEvolution}).
However, the field can vary between repeated experimental runs. This
inevitably leads to different evolutions of the atomic state and therefore
the observed average state populations are washed out. This dephasing
can be modeled by first calculating the temporal evolution of the
considered state $|\Psi\rangle$ in a constant effective magnetic
field $\vec{B}_{eff}$ according to Eq. (\ref{eq:stateEvolution}),
and incoherently averaging over states resulting from the distribution
of different magnetic fields corresponding to different experimental
runs.

In our spin-precession experiments we start with an initial state
$|\Psi\rangle$ and let it precess for a time $t$ giving $|\Psi\rangle(t)$.
Then the population of a chosen analysis state $|\Psi_{a}\rangle$
is obtained from the overlap with the precessed state, averaged over
all possible evolutions. It is given by 

\begin{multline}
P(\Psi_{a})(t)=\\
=\int dB_{x}dB_{y}dB_{z}\left(p_{x}(B_{x})p_{y}(B_{y})p_{z}(B_{z})|\langle\Psi_{a}|\Psi\rangle(t)|^{2}\right),\label{eq:stateAverage}\end{multline}
where the $p_{j}(B_{j})$ are the normalized distributions for the
$j=x,y,z$ components of the effective magnetic field. If, e.g., the
average values are $\bar{B}_{x}=\bar{B}_{y}=\bar{B}_{z}=0$, and the
z-component of the effective magnetic field follows a Gaussian distribution
\begin{equation}
p_{z}(B_{z})=\frac{1}{\sqrt{\pi}\Delta B_{z}}\exp(-(\frac{B_{z}}{\Delta B_{z}})^{2}),\label{eq:magneticNoiseDist}\end{equation}
 then, after initially preparing the atomic spin state $|\psi_{1}\rangle=\frac{1}{\sqrt{2}}(|1,-1\rangle+e^{i\phi}|1,+1\rangle)$,
we get for the probability to stay in this state after a time $t$
\begin{equation}
P(\psi_{1})(t)=\frac{1}{2}\left(1+\exp(-(\frac{t}{T_{2}^{\ast}})^{2})\right).\end{equation}
 Here $T_{2}^{\ast}=\frac{\hbar}{\mu_{B}g_{F}\Delta B_{z}}$ can be
associated with the transverse coherence time of two-level systems
\cite{Bloembergen48}. Note that the decay for this noise model (\ref{eq:magneticNoiseDist})
is not exponential.

As a second example we consider the dephasing of the spin-states $|\psi_{2}\rangle=|1,\pm1\rangle$
in a fluctuating magnetic field along the $x$-axis. By averaging
over the Gaussian distribution $p_{x}(B_{x})$ as in (\ref{eq:magneticNoiseDist})
we find \begin{equation}
P(\psi_{2})(t)=\frac{1}{8}\left(3+5\cdot\exp(-(\frac{t}{T})^{2})\right).\end{equation}
Here, the state population decays with a time constant $T=\frac{2\hbar}{\mu_{B}g_{F}\Delta B_{x}}$,
which is twice as long as $T_{2}^{\ast}$. $P(\psi_{2})(t)$ approaches
the limit of $\frac{3}{8}$, as the corresponding spin evolution leaves
the qubit subspace $\{|1,-1\rangle,|1,+1\rangle\}$. Such a situation
does not occur in two-level systems and represents a more complex
dephasing scenario. For more general cases the integral in (\ref{eq:stateAverage})
is not analytic.

\subsubsection{Effective fluctuation of the light shift\label{sub:DistrLightShifts}}

\begin{figure}
\begin{centering}
\includegraphics{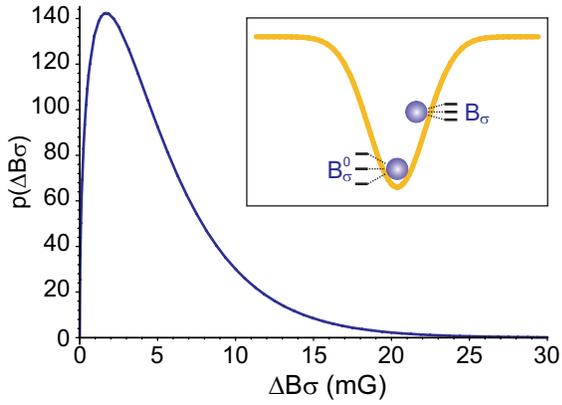} 
\par\end{centering}

\caption{\label{fig:DistrBEff} Probability density $p(\Delta B_{\sigma})$
of the optically induced magnetic field in a three-dimensional harmonic
trap resulting from thermal motion of the atom. The distribution is
plotted as a function of the deviation $\Delta B_{\sigma}=B_{\sigma}^{0}-B_{\sigma}$
of the field $B_{\sigma}$ from its maximal value $B_{\sigma}^{0}$
at the bottom of the trap. The curve was calculated according to Eq.
(\ref{eq:distBsigma}), assuming a trap depth $U_{0}=k_{B}\cdot650\,\mathrm{\mu K}$,
average atomic temperature $T=150\,\mathrm{\mu K}$ and $1\%$ fraction
of circularly polarized trap light. }
\end{figure}

The second part of the effective magnetic field - the vector light
shift - results from the circularly polarized component of the dipole
trap light and is proportional to its intensity. Due to the thermal
motion of the atom in the trap, in each realization of the experiment
the atom will be found at a random position within the trapping potential
and will therefore be subject to a different light shift (see inset
of Fig. \ref{fig:DistrBEff}). Here we shall consider the case, where
the atom can be considered static within one experimental run. In
our experiment the shortest oscillation period in the trap is $44\,\mathrm{\mu s}$,
therefore this assumption is strictly valid only for shorter time
scales. Nevertheless, even for longer experimental times it represents
a worst-case assumption, since a static field which changes from experiment
to experiment leads to a stronger dephasing than a field of the same
amplitude which fluctuates on the experimental time scale or faster
(see Sec. \ref{sub:FieldFluctuations}).

The distribution of positions depends on the thermal energy of the
trapped atom and maps directly onto a distribution of the induced
magnetic field $B_{\sigma}$. To derive this distribution one has
to know the 3D distribution of the corresponding potential energy.
For a thermal energy sufficiently lower than the trap depth the potential
can be considered harmonic and the distribution $p(\Delta B_{\sigma})$
is calculated as follows.

The potential energy $U$ of a 1-dimensional harmonic oscillator can
be written as \begin{equation}
U(E,\varphi)=E\sin(\varphi)^{2},\end{equation}
 where $E$ is the fixed total energy of the motion and $\varphi$
is the phase of the oscillation. We define the potential energy $U$
being non-negative, with $U=0$ at the bottom of the harmonic trap.
If at a certain random moment of time the potential energy is measured,
some random realization for the phase $\varphi\in[0..2\pi]$ will
be found, where every value of $\varphi$ has equal probability. Then
the probability to obtain a value of the potential energy within the
interval $[U,U+dU]$ is $p_{E}(U)dU\propto d\varphi(U)$. Here, $p_{E}(U)$
is the probability density for a given total energy $E$, given by
\begin{equation}
p_{E}(U)=\frac{2}{\pi}\frac{d\varphi}{dU}=\frac{1}{\pi}\frac{1}{\sqrt{U(E-U)}}.\end{equation}
 In thermal equilibrium the total energy $E$ follows a Boltzmann
distribution. The corresponding thermal distribution of the potential
energy $U$ is given by \begin{multline}
p_{1D}(U)=\int_{0}^{\infty}dEp_{E}(U)\frac{1}{k_{B}T}\exp(-\frac{E}{k_{B}T})\\
=\frac{1}{\sqrt{\pi}\sqrt{k_{B}T}}\frac{1}{\sqrt{U}}\exp(-\frac{U}{k_{B}T}).\label{eq:harmOscBoltzmann1D}\end{multline}
For three dimensions the thermal distribution of $U$ is obtained
from a convolution of the three independent 1D distributions \cite{Rosenfeld08_PhD},
resulting in \begin{equation}
p_{3D}(U)=\frac{2}{\sqrt{\pi}(k_{B}T)^{3/2}}\sqrt{U}\exp(-\frac{U}{k_{B}T}).\label{eq:harmOscBoltzmann3D}\end{equation}
The corresponding kinetic energy $E_{kin}$ follows the same distribution.

Here, it is worth mentioning that the distribution (\ref{eq:harmOscBoltzmann3D})
differs from the well-known Maxwell-Boltzmann distribution, which
is $p_{MB}(E)=\frac{1}{2(k_{B}T)^{3}}E^{2}\exp(-\frac{E}{k_{B}T})$.
According to the Virial theorem the average of the potential energy
is half of the total energy: $\left\langle U\right\rangle =\frac{1}{2}E$.
This relation might suggest that the potential and kinetic energy
follow the same distributions as the total energy $E$. However, that
is not the case. The Virial theorem makes only a statement about average
values, while the considered distribution describes the probability
to find a certain value of potential energy at a randomly chosen point
in time (and is therefore not ergodic). Instead, it can be easily
verified that the convolution of the distribution (\ref{eq:harmOscBoltzmann3D})
of the potential energy $U$ and of the identical distribution $p_{3D}(E_{kin})$
of the kinetic energy $E_{kin}$ gives the expected Maxwell-Boltzmann
distribution of the total energy \begin{equation}
\int p_{3D}(U=E-E_{kin})p_{3D}(E_{kin})dE_{kin}=p_{MB}(E).\end{equation}

Now we are able to derive the distribution $p(\Delta B_{\sigma},T)$
of the optically induced magnetic field $B_{\sigma}$ for a thermal
atom at a given temperature $T$. For convenience we introduce the
maximal value of the optically induced magnetic field at the bottom
of the trap $B_{\sigma}^{0}$ according to Eq. (\ref{eq:effectiveField}).
For our typical trap parameters $1\%$ of circular admixture in the
polarization of the trapping light results in $B_{\sigma}^{0}\approx10\,\mathrm{mG}$.
The relation between the induced magnetic field and the potential
energy $U$ is $B_{\sigma}(U)=B_{\sigma}^{0}\frac{U_{0}-U}{U_{0}}$.
Finally, we define the deviation $\Delta B_{\sigma}=B_{\sigma}^{0}-B_{\sigma}$
from the maximal value $B_{\sigma}^{0}$ at the trap center (Fig.
\ref{fig:DistrBEff}, inset). Using these relations we obtain the
distribution of the optically induced magnetic field for an atom at
a temperature $T$ \begin{equation}
p(\Delta B_{\sigma},T)=\frac{2}{\sqrt{\pi}(\frac{B_{\sigma}^{0}}{U_{0}})^{3/2}(k_{B}T)^{3/2}}\sqrt{\Delta B_{\sigma}}\exp(-\frac{\frac{U_{0}}{B_{\sigma}^{0}}\Delta B_{\sigma}}{k_{B}T}).\label{eq:distBsigma}\end{equation}
 This distribution is shown in Fig. \ref{fig:DistrBEff} for typical
experimental parameters.

Expression (\ref{eq:distBsigma}) directly relates the thermal motion
of the trapped atom and the circular admixture in the polarization
of the trapping light to a fluctuating effective magnetic field. These
fluctuations can have a serious impact on the achievable coherence
times as will be shown in Sec. \ref{sec:Experiment}.

\section{Experimental analysis of dephasing mechanisms\label{sec:Experiment}}

\subsection{Single atom trap}

In our experiment a single $^{87}$Rb atom is stored in the optical
dipole trap \cite{APE06,Weber06}, which is loaded from a laser-cooled
cloud of $10^{3}..10^{4}$ atoms of a shallow magneto-optical trap
(MOT). The conservative optical trapping potential is created by a
focused Gaussian laser beam (waist $w_{0}=3.5\,\mathrm{\mu m}$; Rayleigh
range $z_{R}=45\,\mathrm{\mu m}$) at a wavelength of $856\,\mathrm{nm}$,
thereby detuned far to the red of any atomic transition from the atomic
ground level. For a typical power of $P=30\,\mathrm{mW}$ we achieve
a potential depth of $U_{0}=k_{B}\cdot650\,\mathrm{\mu K}$, corresponding
to radial and axial trap frequencies of $\omega_{r}=2\pi\cdot22.7\,\mathrm{kHz}$
and $\omega_{z}=2\pi\cdot1.25\,\mathrm{kHz}$, respectively. This
trap provides a storage time of several seconds, mainly limited by
collisions with the thermal background gas \cite{Weber06}. The fluorescence
light of the trapped atom is collected in a confocal arrangement by
an objective, coupled into a single mode optical fiber and guided
to two single photon counting avalanche photo-detectors (APDs) allowing
also a polarization analysis of single photons. The presence of a
single trapped atom is inferred by detecting fluorescence light. The
bare detection efficiency for a single photon emitted by the atom
is $2\cdot10^{-3}$, including coupling losses into the single-mode
optical fiber and the limited quantum efficiency of the single photon
detectors.

\begin{figure}[t]

\begin{centering}
\includegraphics{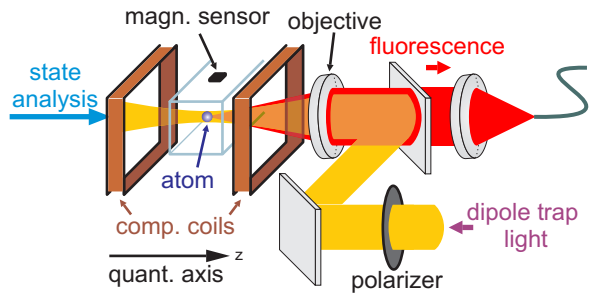} 
\par\end{centering}

\caption{\label{fig:Setup}(Color online) Schematic of the experimental setup
(not to scale). The dipole trap beam is focused into a vacuum cell
by means of a high NA objective. The same objective collects the light
emitted by the atom into a single mode optical fiber. The magnetic
field sensor positioned on the cell surface is used to give a feedback
signal to compensation coils for field stabilization. For simplicity,
only one pair of compensation coils is shown.}
\end{figure}

\subsection{Active magnetic field control\label{sub:magneticFieldControl}}

A crucial requirement for achieving long atomic coherence times is
a precise control of the magnetic field in the region of the optical
dipole trap. For this purpose the fields in our experiment are actively
stabilized. The magnetic field is continuously monitored by a three-axis
magneto-resistive sensor (Honeywell HMC1053), which is located outside
the vacuum glass cell at a distance of $25\,\mathrm{mm}$ from the
position of the trapped atom (see Fig. \ref{fig:Setup}). On a short
time scale the precision of this sensor is limited by electronic noise
(typically $\leq0.1\,\mathrm{mG}$ rms within the effective bandwidth
of $60\,\mathrm{kHz}$). A more significant problem is saturation
of the sensor by the strong fields during loading of the MOT. In this
case a re-magnetization of the sensor is required which limits the
precision to $0.5..1\,\mathrm{mG}$ on long time scales. Using this
sensor we have measured the fluctuations of the magnetic fields and
could identify two main sources. The largest part of the fluctuations
is due to currents drawn by the Munich underground train line passing
at a distance of about $60\,\mathrm{m}$ from our laboratory. The
time scale of these fluctuations is $30\,\mathrm{s}..1\,\mathrm{min}$
with a peak-to-peak amplitude of $20..25\,\mathrm{mG}$ on the vertical
and $6..8\,\mathrm{mG}$ on the horizontal axis. The second major
contribution arises from the $50\,\mathrm{Hz}$ mains supply producing
fluctuations of about $1..2\,\mathrm{mG}$ peak to peak on each axis.
For frequencies higher than the fourth harmonic of the power line
frequency ($200\,\mathrm{Hz}$) the fluctuations were found to be
on the order of $\leq0.3\,\mathrm{mG}$ rms.

The signal from the magnetic field sensor is fed back to compensation
coils by means of a servo loop. The integration time constant was
set such that an active bandwidth of about $200\,\mathrm{Hz}$ was
reached, sufficient to suppress the effects of underground trains
and the power supply line. Given the fluctuations of external fields
described above, our active magnetic field stabilization achieves
an rms stability of $(0.92,\,0.77,\,0.83)\,\mathrm{mG}$ for the three
axes, including re-magnetization precision of the sensor, crosstalk
between different axes $(\leq3.5\%)$ and magnetic field gradients
between the position of the trapped atom and the position of the sensor
$(\leq5\%)$.

\subsection{State preparation and detection}

To study dephasing of a single atomic spin $5^{2}S_{1/2},\, F=1$
we initialize the atom with high fidelity in a well-defined state
of our choice. This is realized by first entangling the atomic Zeeman-states
$|F=1,m_{F}=-1\rangle$ and $|F=1,m_{F}=+1\rangle$ with the polarization
states $|\sigma^{+}\rangle$ and $|\sigma^{-}\rangle$ of a single
emitted photon \cite{APE06,Rosenfeld07,Rosenfeld08} - the entangled
state reads $\left|\Psi^{+}\right\rangle =\frac{1}{\sqrt{2}}(\left|1,-1\right\rangle \left|\sigma^{+}\right\rangle +\left|1,+1\right\rangle \left|\sigma^{-}\right\rangle )$
- and then projecting the atom onto the desired spin-state via a polarization
measurement of the photon. Thus, a measurement of the photon in, e.g.,
the $\sigma^{\pm}$-basis (circularly polarized) leaves the atom in
the $\left|1,\mp1\right\rangle $ state and a measurement in $H/V$-basis
(horizontally/vertically polarized) projects the atom into the $\frac{1}{\sqrt{2}}(\left|1,-1\right\rangle \pm\left|1,+1\right\rangle )$
states, respectively.

After preparation, the atomic spin freely evolves for a defined period
of time in the applied magnetic field $B$. During this time all lasers
except for the one used for the dipole trap are switched off and the
magnetic field is stabilized to a preselected value. Finally, after
a given time, the atomic state detection procedure \cite{APE06} is
applied, allowing us to determine the projection of the atomic state
on any superposition in the $\{\left|1,-1\right\rangle ,\left|1,+1\right\rangle \}$
subspace. By these means we can directly observe the temporal evolution
of selected atomic states (see Fig. \ref{fig:coherence03}).

\subsection{Analysis of the state evolution}

\begin{figure}[t]

\begin{centering}
\includegraphics{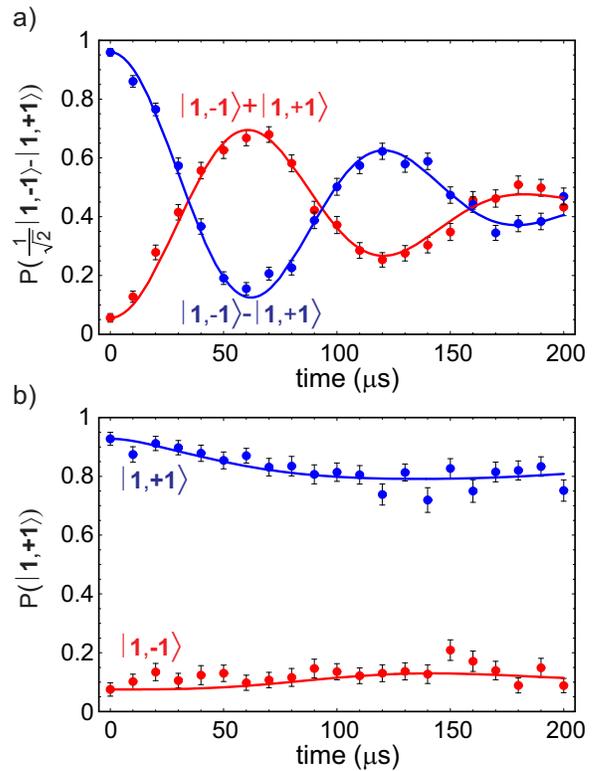} 
\par\end{centering}

\caption{\label{fig:coherence03}(Color online) Temporal evolution of different
atomic spin states. a) Evolution of the superposition states $\frac{1}{\sqrt{2}}(\left|1,-1\right\rangle \pm\left|1,+1\right\rangle )$
in an effective magnetic field of $5.5\,\mathrm{mG}$ along the quantization
axis. b) Evolution of states $\left|1,\pm1\right\rangle $ in a field
compensated to $B\leq2\,\mathrm{mG}$. Measured were the populations
of the spin states $\frac{1}{\sqrt{2}}(\left|1,-1\right\rangle -\left|1,+1\right\rangle )$
in a) and $\left|1,+1\right\rangle $ in b), respectively. The solid
lines represent numerical fits of the measured data to the dephasing
model in Eq. (\ref{eq:stateAverage}), which mainly incorporates fluctuations
of the effective magnetic field due to the vector light shift.}
\end{figure}

In a first measurement, the spin states $\frac{1}{\sqrt{2}}(\left|1,-1\right\rangle \pm\left|1,+1\right\rangle )$
were prepared and a small guiding field of $5.5\,\mathrm{mG}$ was
applied along the quantization axis $z$ such that a slow oscillation
($\sim8\,\mathrm{kHz}$) could be observed. The remaining field components
$B_{x}$ and $B_{y}$ were compensated below $1\,\mathrm{mG}$. After
the prepared spin-states evolved a programmable period of time, the
population of the $\frac{1}{\sqrt{2}}(\left|1,-1\right\rangle -\left|1,+1\right\rangle )$
state was measured, see Fig. \ref{fig:coherence03} (a). For such
a field configuration, where $B_{z}$ dominates all other fields,
the atomic state will stay within the subspace $\{\left|1,-1\right\rangle ,\left|1,+1\right\rangle \}$
during the Larmor precession, as $\left|1,\pm1\right\rangle $ are
eigenstates of the interaction Hamiltonian. We observe the expected
precession of an effective spin-1/2 system with a $1/e$ dephasing
time of about $120\,\mathrm{\mu s}$. In order to extract the parameters
responsible for the dephasing we have numerically fitted the dephasing
model from Eq. (\ref{eq:stateAverage}) to the data points in Fig.
\ref{fig:coherence03} (a). This model includes fluctuations of the
effective magnetic field consisting of residual fluctuations of external
magnetic fields along the $x$-axis (uniformly distributed) and the
dominating distribution $p(\Delta B_{\sigma},T)$ of the optically
induced effective field (\ref{eq:distBsigma}). For the fit we assumed
a trap depth of $U_{0}=650\,\mathrm{\mu K}$ and an average atomic
temperature of $150\,\mathrm{\mu K}$. The Larmor frequency deduced
from this measurement corresponds to an average effective magnetic
field component of $\bar{B}_{z}=(5.5\pm0.5)\,\mathrm{mG}$ composed
of the optically induced field and the externally applied magnetic
field. The observed dephasing is compatible with a standard deviation
of the field distribution of $2.25\,\mathrm{mG}$ along the quantization
axis. From this value we deduce a fraction of $0.6\%$ of circularly
polarized trapping light. This non-negligible fraction is due to the
birefringence of the UHV glass cell where the experiment is performed.
The birefringence is induced by mechanical stress and is not uniform
over the walls of the cell, limiting the degree of control of light
polarization at the position of the atomic trap.

In a second measurement the evolution of the spin states $\left|1,\pm1\right\rangle $
was investigated. Here the absolute value of the effective magnetic
field was compensated to $B\leq2\,\mathrm{mG}$. According to Fig.
\ref{fig:coherence03}(b), the stability of these states largely exceeds
those of superpositions. This can be easily understood since the states
$\left|1,\pm1\right\rangle $ are eigenstates of the effective magnetic
field pointing along the quantization axis $z$, and therefore are
not affected by the fluctuations along this axis. The slower dephasing
of the states $\left|1,\pm1\right\rangle $ shows that fluctuations
along the $x$ and $y$ axes are smaller than along $z$.

\begin{figure*}[t]
\begin{centering}
\includegraphics{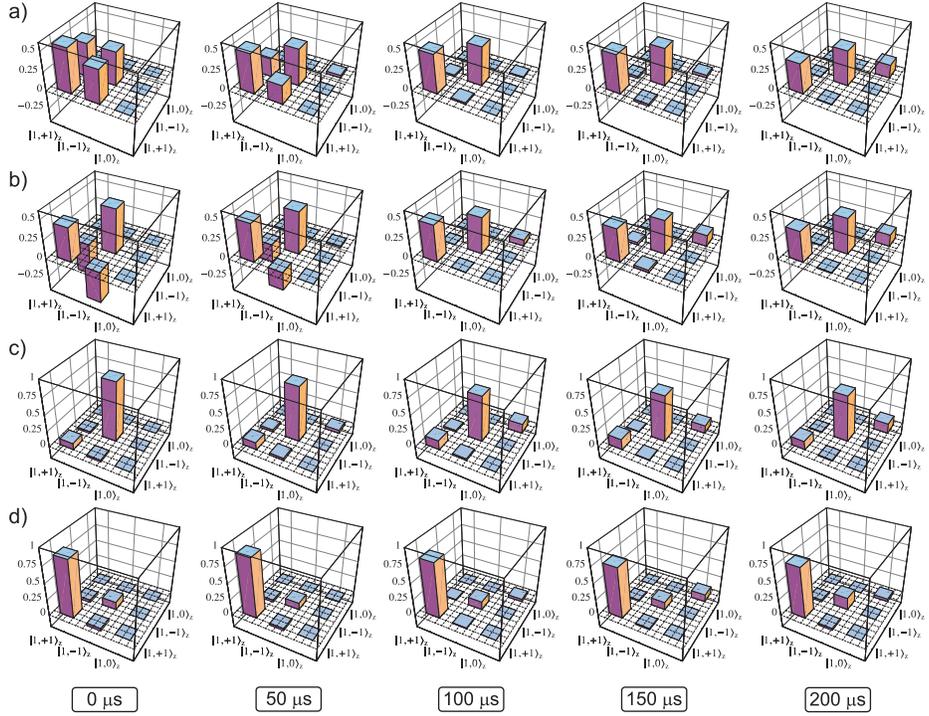} 
\par\end{centering}

\caption{\label{fig:tomographyDM}(Color online) Partial tomographic reconstruction
of the quantum state evolution. Shown are the density matrices (real
part) for the prepared states $\frac{1}{\sqrt{2}}(\left|1,-1\right\rangle \pm\left|1,+1\right\rangle )$
in a) and b), and $\left|1,\mp1\right\rangle $ in c) and d).}
\end{figure*}

\subsection{Quantum state tomography}

The measurement of the temporal evolution of the atomic state provides
a good way to determine its coherence properties. However, state analysis
in one basis does not give complete information about the qubit state
under study. As the ground state $5^{2}S_{1/2},F=1$ of $^{87}$Rb
has total angular momentum of one, the respective temporal evolution
involves three Zeeman sublevels: $m_{F}=\pm1$ and $m_{F}=0$. Thus
the analysis of dephasing processes becomes even more complex compared
to a simple qubit state.

The best way is a complete quantum state tomography (e.g. \cite{quantumTomography}
and references therein), allowing to extract the information on how
the state becomes mixed and also to distinguish coherent and incoherent
processes. Unfortunately, a complete tomography of the spin-1 space
in general requires 5 Stern-Gerlach-like measurements (each providing
the populations of the 3 spin-1 eigenstates along a certain direction)
\cite{spinTomography} which are not accessible in our experiment.
In particular, the coherences between the $\left|1,\pm1\right\rangle $
Zeeman states and the $\left|1,0\right\rangle $ state can not be
measured with our current technique, as the applied stimulated Raman
adiabatic passage pulses analyze only the effective spin-1/2 subspace
$\{\left|1,-1\right\rangle ,\left|1,+1\right\rangle \}$ in a complete
way \cite{APE06,Rosenfeld07}. However, as the detection efficiency
is close to unity, we can infer the population $\rho_{00}$ of the
$|1,0\rangle$ state as the population missing in the $\{\left|1,-1\right\rangle ,\left|1,+1\right\rangle \}$
subspace.

To reconstruct the density matrix $\rho$ of the spin-1 ground state
$5^{2}S_{1/2},F=1$, we use the worst-case assumption that there is
no coherence between the $|1,0\rangle$ state and the $\left|1,\pm1\right\rangle $
states. We therefore set the corresponding off-diagonal density matrix
elements to zero. The corresponding full $3\times3$ spin-1 density
matrix is then given by \begin{equation}
\rho=\left(\begin{array}{cc}
\rho_{s} & \begin{array}{c}
0\\
0\end{array}\\
\begin{array}{cc}
0 & 0\end{array} & \rho_{00}\end{array}\right).\label{ch4-eq:DMmin}\end{equation}
 Here $\rho_{s}$ is the $2\times2$ density matrix of the spin-1/2
subspace $\{\left|1,-1\right\rangle ,\left|1,+1\right\rangle \}$
and $\rho_{00}=1-trace(\rho_{s})$. As we typically measure populations
of the states $\left|1,\pm1\right\rangle $, $\frac{1}{\sqrt{2}}(\left|1,-1\right\rangle \pm\left|1,+1\right\rangle )$
and $\frac{1}{\sqrt{2}}(\left|1,-1\right\rangle \pm i\left|1,+1\right\rangle )$
which are eigenstates of Pauli spin operators $\hat{\sigma}_{z},\hat{\sigma}_{x},$
and $\hat{\sigma}_{y}$, the reduced density matrix $\rho_{s}$ is
fully accessible by our experimental techniques. By combining these
complementary Stern-Gerlach measurements we are able to reconstruct
the spin-1/2 density matrix $\rho_{s}$ given by \begin{equation}
\rho_{s}=\frac{1}{2}\left(\hat{1}+\left\langle \hat{\sigma}_{x}\right\rangle \cdot\hat{\sigma}_{x}+\left\langle \hat{\sigma}_{y}\right\rangle \cdot\hat{\sigma}_{y}+\left\langle \hat{\sigma}_{z}\right\rangle \cdot\hat{\sigma}_{z}\right).\label{eq:partialDM}\end{equation}

Now, in order to find a quantitative measure for the coherent fraction
of the general spin-1 density matrix $\rho$, we decompose $\rho$
as \begin{equation}
\rho=r|\chi\rangle\langle\chi|+(1-r)\frac{1}{3}\hat{1},\end{equation}
 where $|\chi\rangle\langle\chi|$ is the density matrix of the closest
pure state (which can be in general unknown), and $\frac{1}{3}\hat{1}$
represents a completely mixed spin-1 state. The corresponding \emph{purity
parameter} $r$ is the overlap with the closest pure state $|\chi\rangle$
and therefore represents an ideal measure for the coherence of the
investigated state. It can be calculated from the trace of $\rho^{2}$
as \begin{equation}
r=\sqrt{\frac{1}{2}(3\cdot trace(\rho^{2})-1)}.\label{eq:purityParMin}\end{equation}
 For $r=0$ the state under investigation is completely mixed, while
for $r=1$ it is a pure state.

Based on the above procedure, tomographic measurements for the temporal
evolution of spin-1 density matrices $\rho$ for the initial states
$\frac{1}{\sqrt{2}}\left(\left|1,-1\right\rangle \pm\left|1,+1\right\rangle \right)$
and $\left|1,\pm1\right\rangle $ were performed (Fig. \ref{fig:tomographyDM}).
The external magnetic field was set such that as little as possible
Larmor precession could be observed up to $200\,\mathrm{\mu s}$,
and the circular fraction of the dipole light polarization was $\lesssim1\%$.
In the time evolution of the density matrices $\rho$ of the initial
states $\frac{1}{\sqrt{2}}\left(\left|1,-1\right\rangle \pm\left|1,+1\right\rangle \right)$
one can observe several important features. The first one is the decay
of the off-diagonal elements (coherences) as a general sign of dephasing.
Second, a residual Larmor precession can be identified as the off-diagonal
density matrix elements become imaginary (not shown in Fig. \ref{fig:tomographyDM})
and undergo a change of sign. Third, the population of the $\left|1,0\right\rangle $
state continuously increases reaching $\sim15\%$ after $200\,\mathrm{\mu s}$,
that is the qubit subspace is gradually depopulated. In contrast,
for the initial states $\left|1,\pm1\right\rangle $ the major process
during the evolution is only a slowly increasing population of the
$\left|1,0\right\rangle $ state.

\begin{figure}[t]

\begin{centering}
\includegraphics{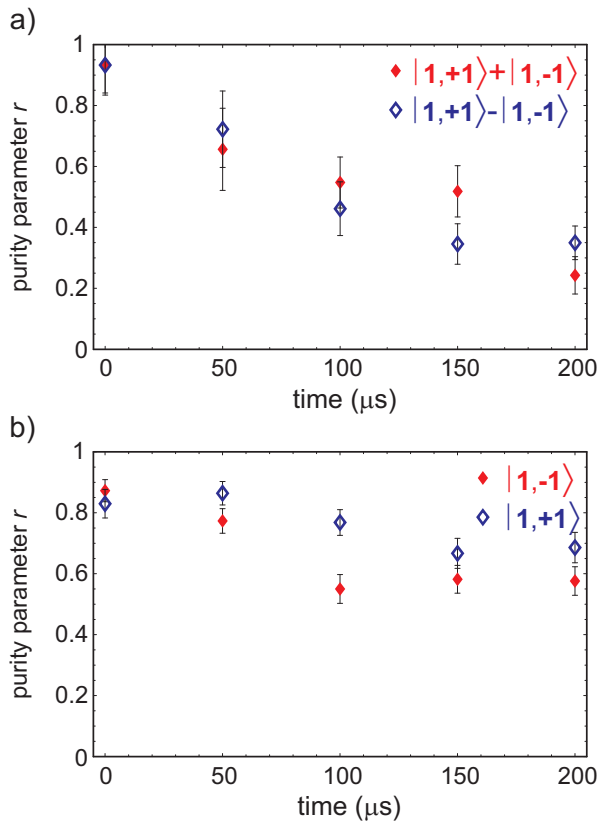} 
\par\end{centering}

\caption{\label{fig:tomographyPurity}(Color online) Purity parameter resulting
from the partial state tomography giving a lower bound for the state
purity \cite{Rosenfeld08}.}
\end{figure}

In order to estimate the coherent fraction of the reconstructed spin-1
density matrices in Fig. \ref{fig:tomographyDM} the corresponding
purity parameter $r$ was evaluated according to Eq. (\ref{eq:purityParMin}),
assessing a lower bound of the atomic spin-coherence. For the evolution
of the superposition states $\frac{1}{\sqrt{2}}\left(\left|1,-1\right\rangle \pm\left|1,+1\right\rangle \right)$,
see Fig. \ref{fig:tomographyPurity}(a), we determine a $1/e$ dephasing
time of $150\,\mathrm{\mu s}$. For the states $\left|1,\pm1\right\rangle $
we estimate the longitudinal dephasing time in absence of a guiding
field by extrapolation to $T_{1}\gtrsim500\,\mathrm{\mu s}$ (see
Fig. \ref{fig:tomographyPurity}(b)). This value gives the time scale
on which the populations of the effective spin-1/2 states $|1,-1\rangle$
and $|1,+1\rangle$ approach an equal mixture of all three spin-1
basis-states $|1,-1\rangle$, $|1,0\rangle$ and $|1,+1\rangle$.

In essence, the significantly longer longitudinal dephasing time shows
that the fluctuations of the effective magnetic field are mainly along
the quantization axis $z$. These fluctuations arise predominantly
from the slow thermal motion of the atom in the trap where a residual
circular polarization admixture leads to a position-dependent vector
light-shift. The resulting dephasing leads to a decay of the off-diagonal
elements of the effective spin-1/2 density matrix $\rho_{s}$ with
a $1/e$ time-constant of $T_{2}^{*}=75\,\mathrm{\mu s}$. The drift
into the $\left|1,0\right\rangle $ state due to magnetic fields orthogonal
to the quantization axis is significantly slower.

\section{Conclusion\label{sec:Conclusion}}

In this paper we have studied the coherence properties of a qubit
encoded in Zeeman substates of the hyperfine ground level of a single
trapped $^{87}$Rb atom. While the {}``fundamental'' decoherence
by Raman scattering of the photons from the dipole trap beam is negligible,
the atomic state can dephase due to technical limitations. The main
mechanisms leading to dephasing were identified as the fluctuations
of stray magnetic fields and the effective magnetic field induced
by the circularly polarized component of the trapping light. By analyzing
the motion of the atom in the trap we have deduced the relation between
the atomic temperature and the fluctuation of the effective magnetic
field due to the circular admixture in the polarization of the trapping
light. The dephasing of atomic memory states was then minimized by
active stabilization of the external magnetic field together with
an accurate setting of the polarization of the dipole trap light.

By performing a partial state tomography of the $5^{2}S_{1/2},\, F=1$
hyperfine ground level we have analyzed the dephasing of different
states. The superposition states like $\frac{1}{\sqrt{2}}(\left|1,-1\right\rangle \pm\left|1,+1\right\rangle )$
show a dephasing time of $75..150\,\mathrm{\mu s}$ which is mainly
limited by field fluctuations along the quantization axis. The spin
states $\left|1,\pm1\right\rangle $ are not sensitive to these fluctuations
and thus show significantly longer dephasing times. Here we want to
stress again that these dephasing times were measured at a magnetic
field close to zero. An externally applied guiding field would induce
a controlled precession of the state while suppressing the influence
of fluctuations orthogonal to its axis. However, the lifting of the
degeneracy of the atomic states coming along with such guiding field
may reduce the fidelity of the atom-photon entanglement scheme. Additionally
it also would require a synchronization of the experiment (in particular
of the time period between preparation and measurement of atomic states)
to the precession period.

In order to further extend the coherence time, two ways for improvement
can be envisaged. On the one hand, better stability of the magnetic
field can be reached by enlarging the geometry of stabilization coils,
thereby reducing field gradients. These measures may be combined with
passive magnetic shielding for better suppression of external magnetic
fields. On the other hand, a large contribution to the dephasing of
the atomic ground state $5^{2}S_{1/2},\, F=1$ results from thermal
motion of the trapped atom in the state-dependent potential induced
by the residual fraction of circularly polarized dipole trap light
($<1\%$). Here longer coherence times could be reached with higher
accuracy of the polarization alignment of the dipole-trap light and
a reduction of birefringende of the glass cell, lowering of the trap
depth, and/or better cooling of the trapped atom. Improvements of
such type will extend the coherence times and thus the usability of
neutral atom quantum memories for future quantum repeater networks.

\section*{Acknowledgments}

This work was supported by the European Commission through the EU
Project Q-ESSENCE and the Elite Network of Bavaria through the excellence
program QCCC.

\end{document}